\documentclass[12pt]{article}
\pdfoutput=1
  \newcommand{\be}[1]{\begin{equation}\label{#1}}
  \newcommand{\ba}[1]{\begin{eqnarray}\label{#1}}

  \newcommand{\rd}{{\rm d}}
  
  \newcommand{\re}{{\rm e}}
  \newcommand{\pa}[1]{\left(#1\right)}
  \newcommand{\paq}[1]{\left[#1\right]}
  
  \newcommand{\av}[1]{\langle#1\rangle}
  \newcommand{\M}{{\rm M_{\rm P}}}

  \def\ee{\end{equation}}
  \def\ea{\end{eqnarray}}

\usepackage{amsmath}

\usepackage{graphicx,latexsym}
\usepackage{graphicx,epsf}
\usepackage{color}
\usepackage{epsfig}
\usepackage{amsmath}
\usepackage[T1]{fontenc}
\usepackage[utf8]{inputenc}
\usepackage{tikz}
\usepackage{authblk}
\begin{document}
%%%%%%%%%%%%%
\title{Born-Oppenheimer meets Wigner-Weyl in Quantum Gravity}
\author[1]{Alexander Y. Kamenshchik\thanks{Alexander.Kamenshchik@bo.infn.it}}
\author[2]{Alessandro Tronconi\thanks{Alessandro.Tronconi@bo.infn.it}}
\author[2]{Giovanni Venturi\thanks{Giovanni.Venturi@bo.infn.it}}
\affil[1]{Dipartimento di Fisica e Astronomia and INFN, Via Irnerio 46,40126 Bologna,
Italy\\

L.D. Landau Institute for Theoretical Physics of the Russian
Academy of Sciences, 119334 Moscow, Russia}
\affil[2]{Dipartimento di Fisica e Astronomia and INFN, Via Irnerio 46, 40126 Bologna,
Italy}

\maketitle

\begin{abstract}
Starting from a Born-Oppenheimer decomposition of the Wheeler-DeWitt equation for the quantum cosmology of the matter-gravity system, we have performed a Wigner-Weyl transformation and obtained equations involving a Wigner function for the scale factor
and its conjugate momentum. This has allowed us to study in more detail than previously the approach to the classical limit of gravitation and the way time emerges in such a limit. To lowest order we reproduce the Friedmann equation and the previously obtained equation for the evolution of matter. We also obtain expressions for higher order corrections to the semi-classical limit.
\end{abstract}

%%%%%%%%%%%%%%%%%%%%%
\section{Introduction}
The success of inflationary cosmology \cite{1} requires an investigation of the physics of the early universe. This leads one to study quantum cosmology, that is the treatment of the universe as a unique quantum object governed by the laws of general relativity and quantum theory. The mathematical structure of quantum gravity
and cosmology is that of a theory with first class constraints, having re-parametrisation invariance or invariance with respect to space-time diffeomorphisms. The main goal of quantum cosmology is the description of the quantum state of the universe. Such a state should satisfy the Wheeler-DeWitt (WDW) equation \cite{2} which results from the application of the Dirac quantisation procedure to the universe. The solutions of the WDW equation involving a limited number of homogeneous degrees of freedom are called mini-superspace models. One may then consider inhomogeneous degrees of freedom on such a homogeneous background. A further consequence of the reparametrization invariance is that in the resulting WDW equation, time (or time derivatives) does not appear. Clearly if one considers as a starting point the universe as a quantum system it must be possible to re-obtain the usual classical Einstein and quantum matter (Schwinger-Tomonaga) equations with, as a boon, quantum gravitational corrections leading to observable (?) effects for sufficiently strong gravitational fields as occur at the beginning of inflation or near black hole horizons. This last step, however, does not appear to be immediate and is closely related to the way the classical limit arises.\\
Depending on the system being considered one has diverse approaches to the classical limit, see e.g. \cite{3} and references therein. One may consider the particle (geometrical optics limit for light) or eikonal limit (WKB) in which, for a path integral representation of the wave function, neighbouring paths will tend to yield cancelling contributions on account of the rapid variation of the phase associated with the exponential of the (effective) action. An exception to this rule occurs at stationary points of the exponent and the associated paths are related to classical trajectories. In another approach the classical limit emerges through the use of the correspondence principle in which one considers large quantum numbers (including the use of coherent states). An example of this is the quantum harmonic oscillator with a large value of the quantum number $n$ (level of the one-dimensional oscillator).\\
A further complication in inflationary cosmology arises because of the presence of matter. As before we shall apply the Born-Oppenheimer approach \cite{4} to the composite quantum matter-gravity system \cite{5,6}. Such an approach is plausible since gravity is characterized by the Planck mass which is much greater than the usual matter mass. This allows one to suitable factorize the wave function of the universe (composite system) into a purely gravitational part in which, to lowest order, gravity is driven by average matter hamiltonian and a matter part wherein, again to lowest order, matter follows gravity adiabatically and time is seen to emerge through a semiclassical approximation to the gravitational wave function.
We wish to again study, in such a context (however see also refs. \cite{6b,6c}), the semiclassical limit in order to examine the emergence of time, however, the method we shall use, in the present paper, is associated with the use of the Wigner function \cite{7}. Such an approach was introduced in order to study quantum corrections to classical statistical mechanics and its goal was to link the wave function to a probability distribution in phase space \cite{8}. Our approach will consist of studying the Wigner-Weyl transformations \cite{9} for the gravitational part of the equations obtained in our BO approach with the aim of studying the semiclassical limit and the associated quantum corrections.\\
The Wigner function was previously introduced in a quantum cosmological
context in \cite{10}. The motivation for this was again to study the possible existence of correlations between variables and their canonical momenta in quantum cosmology and to see how the classical behaviour emerges, thus improving on the use of the simple WKB approximation to the gravitational wave function.  
%%%%%%%%%%%%%%%%%%%%%
\section{Born-Oppenheimer formalism}
We shall consider the inflaton-gravity system which is described by the following action
\be{action}
S=\int d\eta\paq{-\frac{\M^2}{2}a'^2+\frac{a^2}{2}\pa{\phi'^2-2V(\phi) a^2}}+S_{\rm MS}
\ee
where $\M$ is the Planck mass, $a$ is the scale factor, $\eta$ is the conformal time, $\phi$ is the homogenous part of the inflaton, $S_{\rm MS}$ is the action of the Mukhanov-Sasaki field \cite{MS} (which we shall not use in the present manuscript) and we consider a flat Robertson-Walker metric. The hamiltonian for the system then is
\be{ham}
H=-\frac{\pi_a^2}{2\M^2}+\frac{\pi_\phi^2}{2a^2}+a^4V(\phi)+H_{\rm MS}
\ee
where $\pi_a=-M^2 a'$ and $\pi_\phi=a^2\phi'$ and $H_{MS}$ is the Mukhanov-Sasaki contribution which we shall henceforth neglect. The canonical quantisation of the Hamiltonian constraints leads to the following WDW equation for the wave function of the universe (gravity plus matter) 
\be{wdw}
\paq{\frac{\hbar^2}{2\M^2}\frac{\partial^2}{\partial a^2}-\frac{\hbar^2}{2a^2}\frac{\partial^2}{\partial \phi^2}+a^4V(\phi)
}\Psi(a,\phi)\equiv \paq{\frac{\hbar^2}{2\M^2}\frac{\partial^2}{\partial a^2}+\hat H_\phi}\Psi(a,\phi)=0
\ee
which constitutes the starting point for our considerations.\\
Finding the general solution to the WDW equation (\ref{wdw}) even in the absence of perturbations is a very complicated task due to the interaction between matter and gravity. A set of approximate solutions can be found in the BO approach \cite{6b}. The BO approach was originally introduced in order to simplify the Schr\"odinger equation of complex atoms and molecules and has been applied successfully to the inflaton gravity system \cite{6}. It consist of factorizing the wave function $\Psi(a,\phi)$ of the universe into a product 
\be{BO}
\Psi(a,\phi)=\psi(a)\chi(a,\phi),
\ee
where $\chi$ is normalized and not further separable. Coupled equations of motion for $\psi$ and $\chi$ may then be obtained
\be{greq}
\pa{
\frac{\hbar^2}{2\M^2}\partial_a^2+\av{\hat H_\phi}
}\tilde \psi=-\frac{\hbar^2}{2\M^2}\av{\partial_a^2}\tilde \psi,
\ee
\be{mateq}
\frac{\hbar^2}{\M^2}\partial_a\tilde\psi\partial_a\tilde \chi+\tilde \psi\pa{\hat H_\phi-\av{\hat H_{\phi}}
}\tilde \chi=\frac{\hbar^2}{2\M^2}\tilde\psi\paq{\av{\partial_a^2}-\partial_a^2}\tilde\chi,
\ee
where $\av{\hat O}=\langle \tilde \chi|\hat O|\tilde \chi\rangle$, with $\langle \tilde \chi|\tilde \chi\rangle=1$ and
\be{phase}
\psi=\tilde \psi\, \re^{-i\int^a \mathcal {A}\rd a'},\;\chi=\tilde \chi\, \re^{i\int^a \mathcal {A}\rd a'}, \;\mathcal{A}=-i\langle \chi|\partial_a|\chi\rangle.
\ee
The right hand sides (RHS) of Eqs. (\ref{greq}) and (\ref{mateq}) describe non-adiabatic transitions and are generally associated with quantum gravitational effects. \\
In order to discuss the quantum to classical transition for the gravitational sector and correspondingly define the (classical) time evolution for matter, one must consider the solution of the gravity equation and evaluate the probability current associated with it \cite{6b}. This approach has been discussed in several papers \cite{6,6c}. Indeed when the gravity equation cannot be solved exactly, which is often the case, the WKB approach can be used. On considering the $\hbar\rightarrow 0$ limit one then obtains the semiclassical gravitational wave-function (in particular see \cite{6b}) and the Friedmann equation from the gravitational equation (\ref{greq}). On then substituting into the matter equation and keeping the contributions to order $\hbar^2$ one is finally led to a matter equation with quantum gravitational corrections included. Such corrections include the non-adiabatic effects and the quantum fluctuations arising from the introduction of time. However we shall here consider a different approach which consists of introducing the Wigner function for the gravitational sector in order to study how time then emerges. 
%%%%%%%%%%%%%%%%%%%%%%%%%%%%
\subsection{Wigner function for gravity}
The equation for the gravitational wave function is (\ref{greq}). One can define the Wigner function associated with it as
\be{Wdef}
W(a,p)\equiv \int_{-\infty}^{+\infty}\rd s\re^{\frac{i}{\hbar}p s}\tilde \psi_-^{*}\tilde\psi_+,
\ee
where $\tilde\psi_{\pm}=\tilde \psi(a_\pm)$ and $a_\pm=a\pm s/2$. Let us note that the gravitational wave function has support only when its argument is positive and therefore for $a$ fixed $s$ only varies in the interval $\paq{-2a,2a}$. One can, in principle, solve the equation for the gravitational wave function and then calculate the integral (\ref{Wdef}) to obtain the corresponding Wigner function. Apart from few particular cases, this procedure requires a series of approximations to first solve (\ref{greq}) and then evaluate the integral (\ref{Wdef}). Let us note that, in order to evaluate the integral (\ref{Wdef}) one needs an accurate approximation of the gravitational wave function over the whole interval of integration i.e. for values of its argument in $\paq{0,2a}$. In the large or small $a$ limits the gravitational equation often has a simple form and an approximate solution may be easily guessed but this is seldom the case over the whole interval $\paq{0,2a}$ with $a$ large. We shall briefly return to this point in the Conclusions.\\
It appears then more convenient to obtain the exact equations satisfied by (\ref{Wdef}) and afterwards use some approximation scheme to obtain  the Wigner function of the system directly from them.\\
On substituting $a\rightarrow a_+$ into (\ref{greq}), multiplying by $\exp\pa{ips/\hbar}\tilde\psi_-^*$ and integrating over $s$ one has
\be{greq1}
\int_{-\infty}^{+\infty}\rd s\re^{\frac{i}{\hbar}p s}\tilde \psi_-^{*}\paq{\frac{\hbar^2}{2\M^2}\partial_{+}^2+\av{\hat H_\phi}_++\frac{\hbar^2}{2\M^2}\av{\partial_{+}^2}_+
}\tilde \psi_+=0,
\ee
where 
\be{defOpm}
\av{\hat O}_{\pm}\equiv \langle\tilde \chi(a_\pm,\phi)|\hat O| \tilde \chi(a_\pm,\phi)\rangle
\ee
and $\partial_\pm\equiv \partial_{a_\pm}$.\\
Similarly, starting from the equation for $\tilde \psi^*$ with $a\rightarrow a_-$, multiplying by $\exp\pa{ips/\hbar}\tilde\psi_+$ and finally integrating over $s$ one has
\be{greq2}
\int_{-\infty}^{+\infty}\rd s\re^{\frac{i}{\hbar}p s}\tilde \psi_+\paq{\frac{\hbar^2}{2\M^2}\partial_{-}^2+\av{\hat H_\phi}_-+\frac{\hbar^2}{2\M^2}\av{\partial_{-}^2}_-^*
}\tilde \psi_{-}^*=0.
\ee
The resulting equations (\ref{greq1},\ref{greq2}) can now be summed and subtracted obtaining
\ba{greqsd}
&&\int_{-\infty}^{+\infty}\rd s\re^{\frac{i}{\hbar}p s}\paq{\frac{\hbar^2}{2\M^2}\pa{\partial_{+}^2\pm\partial_-^2}+\pa{\av{\hat H_\phi}_+\pm\av{\hat H_\phi}_-}\right.\nonumber\\
&&\left.+\frac{\hbar^2}{2\M^2}\pa{\av{\partial_{+}^2}_+\pm\av{\partial_{-}^2}_-^*}
}\tilde \psi_-^{*}\tilde \psi_+=0
\ea
and we used the fact that $a_+$ and $a_-$ are independent variables. The integrand of (\ref{greqsd}) can be rewritten in terms of $a$ and $s$. Let us note that $\partial_{+}^2+\partial_-^2=(\partial_a^2+4\partial_s^2)/2$ and $\partial_{+}^2-\partial_-^2=2\partial_a\partial_s$. On remembering that the gravitational wave functions are zero on the boundaries, integration by parts leads to
\be{intbp1}
\int_{-\infty}^{+\infty}\rd s\re^{\frac{i}{\hbar}p s} \partial_s^{(n)} \tilde \psi_-^{*}\tilde \psi_+=\pa{-\frac{i}{\hbar}p}^n W(a,p)
\ee 
whereas 
\be{intbp2}
\int_{-\infty}^{+\infty}\rd s\re^{\frac{i}{\hbar}p s} s^{n} \tilde \psi_-^{*}\tilde \psi_+=\pa{-i\hbar}^n \partial_{p}^{(n)}W(a,p).
\ee 
Thus on Taylor expanding the expectation values which appear in (\ref{greqsd}) for $s\rightarrow 0$ and using (\ref{intbp2}) one finally obtains 
\ba{Weq1}
&&\frac{\hbar^2}{4\M^2}\partial_a^2W-\frac{p^2}{\M^2}W+\sum_{n=0}^{\infty}\paq{(-1)^n+1}\frac{(i\hbar)^n}{2^n n!}\frac{\rd^n \av{H_\phi}}{\rd a^n}\partial_{p}^{(n)}W\nonumber\\
&&+\frac{\hbar^2}{2\M^2}\sum_{n=0}^{\infty}\paq{\frac{(-i\hbar)^n}{2^nn!}\frac{\rd^n}{\rd a^n}\av{\partial_a^2}+{\rm c.c.}}\partial_{p}^{(n)}W=0
\ea
and
\ba{Weq2}
&&-\frac{i\hbar\,p}{\M^2}\partial_a W+\sum_{n=0}^{\infty}\paq{(-1)^n-1}\frac{(i\hbar)^n}{2^nn!}\frac{\rd^n\av{H_\phi}}{\rd a^n}\partial_{p}^{(n)}W\nonumber\\
&&+\frac{\hbar^2}{2\M^2}\sum_{n=0}^{\infty}\paq{\frac{(-i\hbar)^n}{2^nn!}\frac{\rd^n}{\rd a^n}\av{\partial_a^2}-{\rm c.c.}}\partial_{p}^{(n)}W=0.
\ea
the latter being a quantum Liouville equation.
The Wigner function must satisfy the above equations simultaneously. Let us note that each equation may contain infinite derivatives w.r.t $p$ unless the Taylor expansion of $\av{H_\phi}$ and $\av{\partial_a^2}$ contains a finite number of contributions. Since each contribution is associated with a different power of $\hbar$ and, if $\av{H_\phi}$ and $\av{\partial_a^2}$ can be written as a finite series of powers of $a$, then higher derivatives with respect to $p$ are multiplied by decreasing coefficients for $\hbar\rightarrow 0$ and $a\rightarrow \infty$.\\
In the classical limit, one must keep the leading contribution for each equation. Equation (\ref{Weq1}) then becomes
\be{Weq1cl}
\paq{-\frac{p^2}{\M^2}+2 \av{H_\phi}}W=0
\ee
which is satisfied by 
\be{Wcl}
W_{\rm cl}(a,p)=\delta\pa{-\frac{p^2}{\M^2}+2 \av{H_\phi}}, 
\ee
i.e. when $p$ is equal to its classical value $p_{\rm cl}=\pm\M\sqrt{2  \av{H_\phi}}$. Equation (\ref{Weq2}), to leading order, is
\be{Weq2cl}
-\frac{i\hbar\,p}{\M^2}\partial_a W-i\hbar\frac{\rd\av{H_\phi}}{\rd a}\partial_{p}W=0
\ee
and is satisfied by any function $W=W(-\frac{p^2}{\M^2}+2 \av{H_\phi})$, in agreement with (\ref{Wcl}).\\
%%%%%%%%%%%%%%%%%%%%%%%%
\subsection{The matter equation}
Given the classical limit for the gravitational wave function one may now study the matter equation (\ref{mateq}). In order to introduce the classical time one must replace $a\rightarrow a_+$, multiply by $\tilde \psi_-^*$ and integrate over $s$. Finally one must integrate over $p$ in an arbitrary integral $\mathcal{I}_{\pm}$ domain around one the two peaks of the Dirac delta (\ref{Wcl}) thus selecting either the contracting or the expanding phase of the Universe.\\
The matter equation then takes the following form
\ba{mateqW}
&&\int_{\mathcal I_{+}} \!\!\!\!\rd p\int\rd s\,\re^{\frac{i}{\hbar}p s}\tilde\psi_-^*\left[\frac{\hbar^2}{\M^2}\pa{\partial_+\tilde\psi_+}\pa{\partial_+\tilde\chi_+}+\tilde\psi_+\pa{H_{\phi,+}-\av{H_{\phi}}_+}\tilde\chi_+\right.\nonumber\\
&&\left.+\frac{\hbar^2}{2\M^2}\tilde\psi_+\pa{\av{\partial_+^2}_+-\partial_+^2}\tilde\chi_+\right]
\ea
where the expanding branch has been chosen. Let us consider the first contribution in (\ref{mateqW}). If one Taylor expands the wave function for matter for $s\rightarrow 0$, then its derivative is
\be{dchi+}
\partial_+\tilde\chi_+=\pa{\frac{1}{2}\partial_a+\partial_s}\pa{\tilde \chi+\partial_a\tilde\chi\frac{s}{2}+\tilde \partial_a^2\chi \frac{s^2}{8}+\dots}=\partial_a\tilde\chi+\partial_a^2\tilde\chi \frac{s}{2}+\dots
\ee
while the derivative of the product of the gravitational wave functions is
\be{dpsipsi}
\partial_+\pa{\tilde\psi_-^*\tilde\psi_+}=\frac{1}{2}\partial_a\pa{\tilde\psi_-^*\tilde\psi_+}+\partial_s\pa{\tilde\psi_-^*\tilde\psi_+}
\ee
On using the relations (\ref{intbp1},\ref{intbp2}) one then finds 
\be{cont1}
\int\rd s\re^{\frac{i}{\hbar}p s}\pa{\partial_a\tilde \chi}\partial_s\pa{\tilde\psi_-^*\tilde\psi_+}=-i\frac{p}{\hbar}\pa{\partial_a\tilde\chi}W\sim \hbar^{-1},
\ee
\be{cont2}
\int\rd s\re^{\frac{i}{\hbar}p s}\pa{\partial_a\tilde \chi}\frac{1}{2}\partial_a\pa{\tilde\psi_-^*\tilde\psi_+}=\frac{1}{2}\pa{\partial_a\tilde\chi}\pa{\partial_a W}\sim \hbar^{0},
\ee
\be{cont3}
\int\rd s\re^{\frac{i}{\hbar}p s}\pa{\partial_a^2\tilde \chi}\frac{1}{2}s\,\partial_s\pa{\tilde\psi_-^*\tilde\psi_+}=-\frac{1}{2}\pa{\partial_a^2\tilde\chi}\pa{W+p\partial_{p}W}\sim \hbar^{0},
\ee
\be{cont4}
\int\rd s\re^{\frac{i}{\hbar}p s}\pa{\partial_a^2\tilde \chi}\frac{1}{2}s\,\partial_a\pa{\tilde\psi_-^*\tilde\psi_+}=\frac{1}{4}\pa{\partial_a^2\tilde\chi}\pa{-i\hbar}\frac{\rd^2 W}{\rd p\rd a} \sim \hbar^{1},
\ee
and higher powers of $s$ in (\ref{dchi+}) contribute higher powers of $\hbar$ to the first contribution in (\ref{mateqW}) and can be neglected.\\
The remaining contributions to Eq. (\ref{mateqW}) have the form
\be{remcon}
\int\rd s\re^{\frac{i}{\hbar}p s}\pa{D_+\tilde \chi_+}\pa{\tilde\psi_-^*\tilde\psi_+}=\pa{D\tilde \chi}W-\partial_a \pa{D\tilde \chi}i\hbar \partial_{p}W+\dots
\ee
where $D$ generically indicates an $a$ dependent operator acting on $\tilde \chi$, $\pa{D_+\tilde \chi_+}$ is expanded in a Taylor series for $s$ small and the ellipsis denote higher orders in $\hbar$. The integral is finally simplified using (\ref{intbp2}) leading to a power series of $\hbar$ and derivative w.r.t. $p$ of the Wigner function.\\
Let us now integrate each contribution with respect to $p$. Since 
\be{Wclsim}
W_{\rm cl}=\delta\pa{-\frac{p^2}{\M^2}+2 \av{H_\phi}}=\frac{\delta\pa{p-\sqrt{2\M^2\av{H_\phi}}}+\delta\pa{p+\sqrt{2\M^2\av{H_\phi}}}}{2\sqrt{2\M^2\av{H_\phi}}}
\ee
Eq. (\ref{cont1}) becomes
\be{cont1b}
-\int_{\mathcal I_{+}} \!\!\!\!\rd pi\frac{p}{\hbar}\pa{\partial_a\tilde\chi}W=-\frac{i}{2\hbar}\partial_a \tilde \chi.
\ee
The contribution (\ref{cont2}) can be simplified by using the classical ``Liouville'' equation (\ref{Weq2cl}) as follows
\ba{cont2b}
&&\int_{\mathcal I_{+}} \!\!\!\!\rd p\frac{1}{2}\pa{\partial_a\tilde\chi}\pa{\partial_a W_{\rm cl}}=-\int_{\mathcal I_{+}} \!\!\!\!\rd p\frac{1}{2}\pa{\partial_a\tilde\chi}\pa{\frac{\rd \av{H_\phi}}{\rd a}\frac{\M^2}{p} \partial_{p}W_{\rm cl}}\nonumber\\
&&=-\left.\partial_a\tilde \chi  \frac{\rd \av{H_\phi}}{\rd a}\frac{\M^2}{2p} W_{\rm cl}\right|_{\mathcal I_+}-\int_{\mathcal I_+}\rd p\partial_a\tilde \chi \frac{\rd \av{H_\phi}}{\rd a}\frac{\M^2}{2p^2} W_{\rm cl}\nonumber\\
&&=-\partial_a\tilde \chi  \frac{\rd \av{H_\phi}}{\rd a}\frac{\M^2}{4p_{\rm cl}^3}=-\partial_a\tilde \chi  \frac{\rd \ln\av{H_\phi}^{1/4}}{\rd a}\frac{1}{2p_{\rm cl}}
\ea
where $p_{\rm cl}=\sqrt{2\M^2\av{H_\phi}}$ and the boundary term vanishes. The third contribution is 
\be{cont3b}
-\int_{\mathcal I_{+}} \!\!\!\!\rd p\frac{1}{2}\pa{\partial_a^2\tilde\chi}\pa{W_{\rm cl}+p\partial_{p}W_{\rm cl}}=-\frac{1}{4p_{\rm cl}}\partial_a^2\tilde \chi\pa{1+\left.p W_{\rm cl}\right|_{\mathcal I_+}-1}=0.
\ee
The fourth contribution is next to next to leading for $\hbar\rightarrow 0$ and is then neglected. Therefore only two of the four terms survive the integration with respect to $p$. The remaining contributions in (\ref{mateqW}) have the form
\be{remconb}
\int_{\mathcal I_{+}} \!\!\!\!\rd p \paq{\pa{D\tilde \chi}W_{\rm cl}-\partial_a \pa{D\tilde \chi}i\hbar \partial_{p}W_{\rm cl}}=\frac{\pa{D\tilde \chi}}{2p_{\rm cl}}-\left.\partial_a \pa{D\tilde \chi}i\hbar W_{\rm cl}\right|_{\mathcal I_+}
\ee
where the last contribution is the boundary term which being proportional to the Dirac delta vanishes.\\
Finally on summing the diverse contributions up to order $\hbar^2$ and multiplying by $2p_{\rm cl}$ one obtains
\be{mateqW3}
-\frac{i\hbar}{\M^2}p_{\rm cl}\partial_a \tilde \chi- \frac{\hbar^2}{\M^2} \frac{\rd \ln\av{H_\phi}^{1/4}}{\rd a}\partial_a\tilde \chi+\pa{\hat H_\phi-\av{\hat H_{\phi}}
}\tilde \chi=\frac{\hbar^2}{2\M^2}\paq{\av{\partial_a^2}-\partial_a^2}\tilde\chi
\ee
and $-\frac{i\hbar}{\M^2}p_{\rm cl}\partial_a \tilde \chi\equiv -i\hbar \partial_\eta$, with $\eta$ the conformal time. We end this Section by noting that (\ref{mateqW3}) is exactly the same as was obtained by the previous approach on using the WKB solution for the gravitational equation. 
%%%%%%%%%%%%%%%%%%%%%%
\section{NLO corrections in $\hbar$}
In the previous Section we found that, to leading order (LO), the correct classical limit (Friedmann) is reproduced for $\hbar\rightarrow 0$. In this Section we are interested in evaluating the next to leading order (NLO) corrections to the classical limit for the gravitational part of the system which will then have consequences on all matter evolution.\\
To the NLO, the Eq. (\ref{Weq1}) is
\be{greqNLO}
\pa{p^2-p_{\rm cl}^2}W=\frac{\hbar^2}{4}\paq{\partial_a^2W-\M^2\frac{\rd^2 \av{H_\phi}}{\rd a^2}\partial_{p}^2W
+2\pa{\av{\partial_a^2}+{\rm c.c.}}W}+\mathcal{O}\pa{\hbar^4}
\ee
and we shall again just consider the positive branch (expanding universe case). When the classical limit ($\hbar\rightarrow 0$) is studied one is interested in calculating the solution of  (\ref{greqNLO}) in a small interval around the classical solution $p/p_{\rm cl}\in \paq{1-\epsilon,1+\epsilon}$ and
\be{diffp}
\pa{p^2-p_{\rm cl}^2}W=\pa{p+p_{\rm cl}}\pa{p-p_{\rm cl}}W= 2 p_{\rm cl}\pa{p-p_{\rm cl}}W+\mathcal{O}\pa{\epsilon^2}
\ee
The quantum Liouville equation (\ref{Weq2}) and its derivative with respect to the scale factor can then be used to simplify Eq. (\ref{greqNLO}). On retaining  contributions up to $\hbar^2$ one finally obtains
\ba{greqNLOb}
&&\frac{\hbar^2}{4}\paq{\frac{1}{2}\frac{\rd^2 p_{\rm cl}^2}{\rd a^2}\pa{\frac{\partial_{p}W}{p_{\rm cl}}+\partial_{p}^2 W}+\pa{\frac{\rd p_{\rm cl}}{\rd a}}^2\pa{\frac{\partial_{p}W}{p_{\rm cl}}-\partial_{p}^2W}}\nonumber\\
&&+\paq{2 p_{\rm cl}\pa{p-p_{\rm cl}}-\frac{\hbar^2}{2}\pa{\av{\partial_a^2}+{\rm c.c.}}}W
=0,
\ea
where we replaced $p\rightarrow p_{\rm cl}$ in the $\mathcal{O}\pa{\hbar^2}$ part.
This last equation (\ref{greqNLOb}) can be solved exactly in momentum space. On setting
\be{defWt}
W(a,p)=\int_{-\infty}^{+\infty} \rd y \re^{i\,y\,p}\widetilde W(a,y),
\ee
one finds the following equation for the transformed Wigner function $\widetilde W$
\ba{greqNLOk}
&&\frac{\hbar^2}{4}\paq{p_{\rm cl}\frac{\rd^2 p_{\rm cl}}{\rd a^2}+2\pa{\frac{\rd p_{\rm cl}}{\rd a}}^2}\frac{iy\widetilde W}{p_{\rm cl}}-\frac{\hbar^2}{4}p_{\rm cl}\frac{\rd^2 p_{\rm cl}}{\rd a^2}y^2 \widetilde W\nonumber\\
&&+2ip_{\rm cl}\frac{\rd\widetilde W}{\rd y}-\paq{2 p_{\rm cl}^2+\frac{\hbar^2}{2}\pa{\av{\partial_a^2}+{\rm c.c.}}}\widetilde W=0
\ea
which can be easily solved obtaining
\ba{solWt}
&&\widetilde W=\widetilde W_0\exp\left\{-\frac{\hbar^2}{8}\frac{\rd^2 p_{\rm cl}}{\rd a^2}\frac{i y^3}{3}-\frac{\hbar^2}{16}\paq{\frac{\rd^2 p_{\rm cl}}{\rd a^2}+\frac{2}{p_{\rm cl}}\pa{\frac{\rd p_{\rm cl}}{\rd a}}^2}\frac{y^2}{p_{\rm cl}}\right.\nonumber\\
&&\left.-\paq{p_{\rm cl}+\frac{\hbar^2}{4p_{\rm cl}}\pa{\av{\partial_a^2}+{\rm c.c.}}}iy\right\}.
\ea
The above expression can be transformed back so as to obtain the Wigner function with the $\mathcal{O}\pa{\hbar^2}$ corrections included. Let us note that the Airy function ${\rm Ai}(x)$ admits an integral representation \cite{14} which, on shifting the integration variable and rescaling it properly, leads to the following relation
\be{Aiint3b}
\int_{-\infty}^{+\infty}\rd t\,\re^{i\pa{\frac{b}{3}t^3+ct^2+dt}}=\frac{2\pi}{b^{1/3}}\,\exp\paq{i \frac{c}{b}\pa{\frac{2c^2}{3b}-d
 }}{\rm Ai}\paq{\frac{d-\frac{c^2}{b}}{b^{1/3}}}
\ee
which can be used in (\ref{defWt}) and (\ref{solWt}) setting
\ba{WtWrels}
b&=&-\frac{\hbar^2}{8}\frac{\rd^2 p_{\rm cl}}{\rd a^2},\\
c&=&\frac{i\hbar^2}{16p_{\rm cl}}\paq{\frac{\rd^2 p_{\rm cl}}{\rd a^2}+\frac{2}{p_{\rm cl}}\pa{\frac{\rd p_{\rm cl}}{\rd a}}^2},\\
d&=&p-p_{\rm cl}-\paq{\frac{\hbar^2}{4p_{\rm cl}}\pa{\av{\partial_a^2}+{\rm c.c.}}}.
\ea
Let us observe that \cite{11}
\be{airylimit}
\frac{1}{|\alpha|}{\rm Ai}\pa{\frac{x}{\alpha}}\stackrel{\alpha\rightarrow 0}{=}\delta(x)+\frac{\alpha^3}{3}\delta^{(3)}(x).
\ee
and to the LO and the $\hbar\rightarrow 0$ limit
\ba{Wclcheck}
&&\!\!\!\!\!\!\!\!\!\!\!\!\!\!\!\!\!\!\!\!W\pa{a,p}\simeq \frac{2\pi}{b^{1/3}}{\rm Ai}\paq{\frac{p-p_{\rm cl}+\mathcal{O}\pa{\hbar^2}}{b^{1/3}}}\nonumber\\
&&\!\!\!\!\!\!\!\!\!\!\!\!\!\!\!\!\!\!\!\!\times \exp\paq{\pa{1+\frac{2}{p_{\rm cl}\frac{\rd^2 p_{\rm cl}}{\rd a^2}}\pa{\frac{\rd p_{\rm cl}}{\rd a}}^2}\frac{\pa{p_{\rm cl}-p}}{2\,p_{\rm cl}}+\mathcal{O}\pa{\hbar^2}}\stackrel{\hbar\rightarrow 0}{=} 2\pi\delta\pa{p-p_{\rm cl}}
\ea
with $b^{1/3}\sim \hbar^{2/3}$ and thus the  correct classical limit is recovered. Let us note that the NLO correction to the limit of the Airy function (\ref{airylimit}) is proportional to $b\sim \hbar^2$ and the other contributions, of the same order, must be added  in order to properly calculate the NLO corrections to the limit of the expression (\ref{Aiint3b}). The above results appear rather unwieldy for an immediate use ( for example in the general matter evolution equation ) and we shall return to their application and significance in the conclusions.
%%%%%%%%%%%%%%%%%%%%%%%%
%%%%%%%%%%%%%%%%%%%%%%%%
%%%%%%%%%%%%%%%%%%%%%%%%
%%%%%%%%%%%%%%%%%%%%%%%%
%%%%%%%%%%%%%%%%%%%%%%%
%%%%%%%%%%%%%%%%%%%%%%%
\section{Conclusions}
Following a Born-Oppenheimer decomposition of the wave function of the Universe separate wave equations for the matter and gravity parts of it were obtained from the WDW equation it satisfies. On then contracting both the resulting equations with respect to the gravitational wave function and performing a Wigner-Weyl transformation, one is led to wave equations involving gravitational phase space. In particular since we limited ourselves to the homogeneous part of the gravitational wave function these are the scale factor and its conjugate momentum.
The resulting wave equations will then involve, besides the matter wave function, the Wigner function for gravitation. This Wigner function is of particular interest since it is only in the classical limit for gravitation that time appears, otherwise it is absent in a quantum formulation (in the current approach matter is allowed to remain in a quantum state, since,presumably, the energies for quantum gravity are much higher than those for quantum matter and therefore gravity becomes classical first in an expanding universe \cite{6}). We then concentrated our attention on the resulting constraints on the gravitational Wigner function.\\
The equation for the gravitational Wigner function was further separated into a part relating the Wigner function to matter and a quantum Liouville equation for it. Both included an infinite expansion with respect to the Planck constant. On taking the classical limit for both the equations, the Friedmann equation is recovered and on substitution of the classical solution for the Wigner function into the equation for quantum matter the previously obtained \cite{6} evolution equation for it (Schr\"odinger or Schwinger-Tomonaga) is reproduced.\\
The equations for the Wigner function were then solved while retaining higher order terms in the Planck constant. In particular we kept terms to order $\hbar^2$ which are of the same order as the non-adiabatic contributions to matter, which we previously considered perturbatively in the context of the Mukhanov equation which our new corrections will also effect. We found an Airy function type solution which on taking the classical limit reduces, in lowest leading order, to the previously obtained one together with corrections which can be expressed in terms of generalised functions.\\
Concerning this last point we observe that \cite{11} such an expansion requires that the Airy function be multiplied by a relatively slowly varying function of its argument, it may well be that this is not always possible, in particular for situations for which the classical limit may not exist. Lastly we observe that the Airy function in Eq. (50) was obtained by studying our equations in the vicinity of the (classical) maximum of the Airy, indeed it falls exponentially for $p>p_{\rm cl}$ while it damps oscillatorily on the $p<p_{\rm cl}$ side. Our corrections to the previous introduction of time appear rather unwieldy thus it is convenient to apply them to the evolution of matter after the choice of a suitable inflationary potential.\\
Finally let us conclude by observing that with respect to previous treatments of the Wigner function in cosmology we have included matter and exhibited higher order corrections in Planck's constant. One point we mentioned in Section 2.1 which has however generally been glossed over is that $a>0$ and the gravitational wave function does not have support for $a<0$. This actually, as we saw, has strong consequences on the domain of integration in the Wigner-Weyl transformation (although for the case of suitably localised gravitational wave functions, for example in the presence of a bounce for a small, one need not be concerned). The detailed implementation of such a constraint is cumbersome and has been attempted \cite{12} and studied \cite{13}. 
%%%%%%%%%%%%%%%%%%%%%%%

\section{Acknowledgements}
Alexander Y. Kamenshchik is supported in part by the Russian Foundation for Basic Research grant No. 20-02-00411.

%%%%%%%%%%%%%%%%%%

%%%%%%%%%%%%%%%%%%%%

\begin{thebibliography}{99}

\bibitem{1}A.A. Starobinsky, Phys. Lett. B 91, 99 (1980); A. H. Guth, Phys. Rev. D 23, 347 (1981); A. D. Linde, Phys. Lett. B 129, 177 (1983); A.A. Starobinsky, in H.J. De Vega and N. Sanchez (eds.) Current trends in field theory quantum gravity and strings, Lecture Notes in Physics 246 Springer Verlag, Heidelberg, 1986, pp. 107- 126; A.D. Linde. Particle Physics and Inflationary Cosmology, Harwood, New York, 1990.

\bibitem{2}B.S. DeWitt, Phys. Rev. {\bf 160}, 113 (1967);

\bibitem{3} E. G. Peter Rowe, European J. Physics 8, 81 (1987);

\bibitem{4}M. Born and J.R. Oppenheimer, Ann. Physik {\bf 84}, 457 (1927); 
C. A. Mead and D. G. Truhlar, J. Chem. Phys. {\bf 70}, 2284 (1979);
C. A. Mead, Chem. Phys {\bf 49}, 23 (1980)
C. A. Mead, Chem. Phys {\bf 49}, 33 (1980

\bibitem{5} R. Brout and G. Venturi, Phys. Rev. D {\bf 39}, 2436 (1989)
 
\bibitem{6}G. Venturi, Class. Quantum Grav. {\bf 7}, 1075 (1990);
F.~Finelli, G.~P.~Vacca and G.~Venturi,
%``Chaotic inflation with a scalar field in nonclassical states,''
Phys. Rev. D \textbf{58} (1998), 103514
doi:10.1103/PhysRevD.58.103514
%[arXiv:gr-qc/9712098 [gr-qc]].
A.~Y.~Kamenshchik, A.~Tronconi and G.~Venturi,
 %``The Born?Oppenheimer method, quantum gravity and matter,''
 Class.\ Quant.\ Grav.\ {\bf 35} (2018) no.1, 015012
 doi:10.1088/1361-6382/aa8fb3
% [arXiv:1709.10361 [gr-qc]]
A.~Y.~Kamenshchik, A.~Tronconi and G.~Venturi,
 %``Inflation and Quantum Gravity in a Born-Oppenheimer Context,''
 Phys.\ Lett.\ B {\bf 726} (2013) 518
 doi:10.1016/j.physletb.2013.08.067
 %[arXiv:1305.6138 [gr-qc]].
 %%CITATION = doi:10.1016/j.physletb.2013.08.067;%%
 %26 citations counted in INSPIRE as of 28 Aug 2019
 D.~Bini, G.~Esposito, C.~Kiefer, M.~Kraemer and F.~Pessina,
%``On the modification of the cosmic microwave background anisotropy spectrum from canonical quantum gravity,''
Phys. Rev. D \textbf{87} (2013) no.10, 104008
doi:10.1103/PhysRevD.87.104008
%[arXiv:1303.0531 [gr-qc]].
 A.~Y.~Kamenshchik, A.~Tronconi and G.~Venturi,
 %``Signatures of quantum gravity in a Born?Oppenheimer context,''
 Phys.\ Lett.\ B {\bf 734} (2014) 72
 doi:10.1016/j.physletb.2014.05.028
% [arXiv:1403.2961 [gr-qc]].
 %%CITATION = doi:10.1016/j.physletb.2014.05.028;%%
 %23 citations counted in INSPIRE as of 28 Aug 2019
 A.~Y.~Kamenshchik, A.~Tronconi and G.~Venturi,
 %``Quantum Gravity and the Large Scale Anomaly,''
 JCAP {\bf 1504} (2015) no.04, 046
 doi:10.1088/1475-7516/2015/04/046
% [arXiv:1501.06404 [gr-qc]].
 %%CITATION = doi:10.1088/1475-7516/2015/04/046;%%
 %13 citations counted in INSPIRE as of 28 Aug 2019
 A.~Y.~Kamenshchik, A.~Tronconi and G.~Venturi,
 %``Quantum Cosmology and the Evolution of Inflationary Spectra,''
 Phys.\ Rev.\ D {\bf 94} (2016) no.12, 123524
 doi:10.1103/PhysRevD.94.123524
% [arXiv:1609.02830 [gr-qc]].
 %%CITATION = doi:10.1103/PhysRevD.94.123524;%%
 %10 citations counted in INSPIRE as of 28 Aug 2019

\bibitem{6b}
A.~Tronconi, G.~P.~Vacca and G.~Venturi,
%``The Inflaton and time in the matter gravity system,''
Phys. Rev. D \textbf{67} (2003), 063517
doi:10.1103/PhysRevD.67.063517
[arXiv:gr-qc/0302030 [gr-qc]].


\bibitem{6c}
 A.~Y.~Kamenshchik, A.~Tronconi, T.~Vardanyan and G.~Venturi,
 %``Time in quantum theory, the Wheeler?DeWitt equation and the Born?Oppenheimer approximation,''
 Int.\ J.\ Mod.\ Phys.\ D {\bf 28} (2019) no.06, 1950073
 doi:10.1142/S0218271819500731
 % [arXiv:1809.08083 [gr-qc]].
 %%CITATION = doi:10.1142/S0218271819500731;%%
 %6 citations counted in INSPIRE as of 25 Nov 2019
A.~Y.~Kamenshchik, A.~Tronconi, T.~Vardanyan and G.~Venturi,
 %``Quantum Gravity, Time, Bounces and Matter,''
 Phys.\ Rev.\ D {\bf 97} (2018) no.12, 123517
 doi:10.1103/PhysRevD.97.123517
 %[arXiv:1804.10075 [gr-qc]].
 %%CITATION = doi:10.1103/PhysRevD.97.123517;%%
 %7 citations counted in INSPIRE as of 28 Aug 2019
A.~Y.~Kamenshchik, A.~Tronconi and G.~Venturi,
%``Quantum cosmology and the inflationary spectra from a nonminimally coupled inflaton,''
Phys. Rev. D \textbf{101} (2020) no.2, 023534
doi:10.1103/PhysRevD.101.023534
[arXiv:1911.10918 [gr-qc]].
A.~Y.~Kamenshchik, A.~Tronconi and G.~Venturi,
%``The Born-Oppenheimer approach to Quantum Cosmology,''
[arXiv:2010.15628 [gr-qc]].

\bibitem{7} E. P. Wigner (1932). "On the quantum correction for thermodynamic equilibrium". Phys. Rev. 40 (5): 749–759.
 
\bibitem{8} For a general review and selected papers see C. Zachos, D. Fairlie, T. Curtright: Quantum mechanics in phase space: an overview with selected papers, World Scientific, 2005 ISBN 978-981-4520-43-0

\bibitem{9}W.B.Case,Am.J.Phys.76 (10), 937 (2008); J.S.Ben-Benjamin, M.B.Kim,W.P.Schleich,W.B.Case,L.Cohen, Fortschr. Phys. 65, No. 6–8, 1600092 (2017) 

\bibitem{10}A.Anderson, Phys.Rev.D 42 (1990) 585; J.J.Halliwell, Phys.Rev.D 36, 3626 (1990); S.Habib and R.Laflamme, Phys.Rev.D 42 (1990)4056; S.Habib,Phys.Rev.D 42, 2566 (1990); H.Kodama, ln Fifth Marcel Grossman Meeting, proceedings, Perth, australia, 1988, edited by D.G.Blair and M.J.Buckingham, (World Scientific,Singapore, 1989);
E.Calzetta and B.L.Hu, Phys.Rev.D 40, 380 (1989)

\bibitem{MS}
V.F. Mukhanov, Sov. Phys. JETP {\bf 68}, 1297 (1988);
J.~M.~Maldacena,
 %``Non-Gaussian features of primordial fluctuations in single field inflationary models,''
 JHEP {\bf 0305} (2003) 013;
 V.~F.~Mukhanov, H.~A.~Feldman and R.~H.~Brandenberger,
 %``Theory of cosmological perturbations. Part 1. Classical perturbations. Part 2. Quantum theory of perturbations. Part 3. Extensions,''
 Phys.\ Rept.\ {\bf 215} (1992) 203.
 V.F. Mukhanov, Phys. Lett. B {\bf 218}, 17 (1989);
 J.~M.~Bardeen,
 %``Gauge Invariant Cosmological Perturbations,''
 Phys.\ Rev.\ D {\bf 22}, 1882 (1980).
 doi:10.1103/PhysRevD.22.1882
 M.~Sasaki,
 %``Gauge Invariant Scalar Perturbations in the New Inflationary Universe,''
 Prog.\ Theor.\ Phys.\ {\bf 70} (1983) 394.
 doi:10.1143/PTP.70.394
 %%CITATION = doi:10.1143/PTP.70.394;%%
 %63 citations counted in INSPIRE as of 21 Nov 2019
 
\bibitem{14} O. Val\'ee, M. Soares: ``Airy functions and applications to physics'' (World Scientific) September 2004 https://doi.org/10.1142/p345

 

\bibitem{11}E.J.Heller, J. Chem. Phys. 68, 2066 (1978), E.A.Gislason, J. Chem. Phys.58, 3702, (1973);

\bibitem{12}R.Cordero, H.Garcia-Compean, F.J.Turrubiates, Phys.Rev. D 83, 125030 (2011);
\bibitem{13}see e.g. N.C.Dias and J.N.Prata J.Math.Phys. 43, 4602 (2002);



\end{thebibliography}
\end{document}